# Unusual mixed valence of Eu in two new materials $EuSr_2Bi_2S_4F_4$ and $Eu_2SrBi_2S_4F_4$: Mössbauer and X-ray photoemission Spectroscopy investigations

Zeba Haque[1], Gohil Singh Thakur[1], Rangasamy Parthasarathy[1], Birgit Gerke[2], Theresa Block[2], Lukas Heletta[2], Rainer Pöttgen[2], Amish G. Joshi[3], Ganesan Kalai Selvan[4], Sonachalam Arumugam[4], Laxmi Chand Gupta[1,†], and Ashok Kumar Ganguli[1,5]*

[1] Department of Chemistry, Indian Institute of Technology, New Delhi 110016, India

[2] Institut für Anorganische und Analytische Chemie, Universität Münster, Corrensstrasse 30, D-48149 Münster, Germany

[3] CSIR-National Physical Laboratory, Dr. K.S. Krishnan Road, New Delhi 110012, India

[4] Centre for High Pressure Research, School of Physics, Bharathidasan University, Tiruchirapalli, India 620024

[5] Institute of Nano Science & Technology, Habitat Centre, Mohali 160062, India

*Email: ashok@chemistry.iitd.ac.in

**Abstract**

We have synthesized two new Eu-based compounds, $EuSr_2Bi_2S_4F_4$ and $Eu_2SrBi_2S_4F_4$ which are derivatives of $Eu_3Bi_2S_4F_4$, an intrinsic superconductor with $T_c$ = 1.5 K. They belong to a tetragonal structure (SG: $I4/mmm$, $Z = 2$), similar to the parent compound $Eu_3Bi_2S_4F_4$. Our structural and $^{151}$Eu Mössbauer spectroscopy studies show that in $EuSr_2Bi_2S_4F_4$, Eu-atoms exclusively occupy the crystallographic $2a$-sites. In $Eu_2SrBi_2S_4F_4$, $2a$-sites are fully occupied by Eu-atoms and the other half of Eu-atoms and Sr-atoms together fully occupy $4e$-sites in a statistical distribution. In both compounds Eu atoms occupying the crystallographic $2a$-sites are in a homogeneous mixed valent state ~ 2.6 - 2.7. From our magnetization studies in an applied H ≤ 9 Tesla, we infer that the valence of Eu-atoms in $Eu_2SrBi_2S_4F_4$ at the $2a$-sites exhibits a shift towards 2+. Our XPS studies corroborate the occurrence of valence fluctuations of Eu and after Ar-ion sputtering show evidence of enhanced population of $Eu^{2+}$- states. Resistivity measurements, down to 2 K suggest a semi-metallic nature for both compounds.

**Introduction**

The recently known BiS$_2$-layer containing superconductors have gained considerable attention. Broadly categorising, these superconducting materials comprise the families Bi-O-S (T$_c$ = 4.5-6 K)[1–5], $Ln$O$_{1-x}$F$_x$BiS$_2$ ($Ln$ = La, Ce, Pr, Nd, Sm and Yb) (T$_c$ = 1.9-5.4 K)[6–12], and Sr$_{1-x}$$Ln$$_x$FBiS$_2$ ($Ln$ = La, Ce, Pr, Nd and Sm) (T$_c$ ~ 3 K)[13–17]. The parent $Ln$OBiS$_2$ ($Ln$ = La, Ce and Th) and SrFBiS$_2$ compounds are semiconducting[18–20]. In these compounds, superconductivity is induced by doping them with suitable electron donors[6,13,16,17]. Application of external hydrostatic pressure[21–25] results in further enhancement of T$_c$. The two Eu-based BiS$_2$ superconductors EuFBiS$_2$ (Eu-1112) (T$_c$ = 0.3 K)[26] and Eu$_3$Bi$_2$S$_4$F$_4$ (Eu-3244) (T$_c$ = 1.5 K)[27] are different from other BiS$_2$-superconductors in that they are intrinsic superconductors, namely, no external doping is required to induce superconductivity. Superconductivity in Eu-1112 and Eu-3244 takes place due to the mixed valence state of europium ions (Eu$^{2+}$⇔ Eu$^{3+}$) which effectively creates electron-doping. Enhancement of T$_c$ of Eu-1112 (T$_c$ = 8.6 K at 1.86 GPa) and Eu-3244 (T$_c$ = 10 K at 2 GPa) has been reported by chemical and external pressure effects[28–31].

The selenium substituted derivative Eu$_3$Bi$_2$S$_{4-x}$Se$_x$F$_4$ (0 ≤ x ≤ 2) of Eu$_3$Bi$_2$S$_4$F$_4$ have been known to show T$_c$ higher than that of Eu$_3$Bi$_2$S$_4$F$_4$ and it also exhibits a negative chemical pressure effect. The layered Eu$_3$Bi$_2$S$_4$F$_4$ is made up of two EuFBiS$_2$ layers linked by a EuF$_2$ layer in between. Here the fluorite type Eu$_3$F$_4$ layers and the rocksalt type BiS$_2$ bilayers alternate along the crystallographic $c$-axis.

We have successfully synthesized two new Eu-based compounds, Eu$_2$SrBi$_2$S$_4$F$_4$ and EuSr$_2$Bi$_2$S$_4$F$_4$, derived from Eu$_3$Bi$_2$S$_4$F$_4$ and studied their structural, electrical and magnetic properties. More specifically, we have investigated the phenomenon of valence fluctuations of europium, using $^{151}$Eu-Mössbauer and XPS techniques, which Eu-ions occupying specific crystallographic 2*a*-sites in these materials exhibit. Here we present the results of our investigations.

**Experimental**

Polycrystalline samples of nominal compositions $Eu_{3-x}Sr_xBi_2S_4F_4$ (x = 1 and 2) were synthesized by heating $EuF_3$, $EuF_2$, SrS, $SrF_2$, $Bi_2S_3$ and Bi metal. $Bi_2S_3$ was pre-synthesized by the reaction of Bi and S at 500 °C, respectively for 12 h in vacuum. Stoichiometric amounts of the reactants were mixed well, pelletized, sealed in evacuated quartz tubes and heated at 850°C for 35 h. The heat-treated pellets were dark black, non-lustrous, and remained stable in air for several weeks. They were, however, preserved in an Ar-filled glove-box to avoid any unnecessary contamination. The phase purity of the samples was checked by powder X-ray diffraction using Cu Kα radiation using a Bruker D8 advance diffractometer. Rietveld refinement analysis of the powder X-ray diffraction data was carried out using the TOPAS software package[32]. Resistivity measurements in the temperature range of 2−300 K was carried out using a conventional four-probe method in a Quantum Design Physical Property Measurement System (QD PPMS) Evercool-II. Magnetization measurements on the polycrystalline samples were carried out in the temperature range of 2-300 K using a PPMS (Quantum Design) with a VSM option.

The 21.53 keV transition of $^{151}$Eu with an activity of 130 MBq (2 % of the total activity of a $^{151}$Sm:EuF$_3$ source) was used for the Mössbauer spectroscopic characterization. The measurements were conducted in transmission geometry with a commercial (78 K and 5 K) cryostat, while the source was kept at room temperature. The $Eu_2SrBi_2S_4F_4$ and $EuSr_2Bi_2S_4F_4$ samples were placed in thin-walled PMMA containers at an optimized thickness. Fitting of the spectra was performed with the Normos-90 program system[33].

X-ray Photoemission Spectroscopy (XPS) experiments were carried out using an Omicron Multi-probe® Surface Science System; equipped with a monochromatic source (XM 1000) and a hemispherical electron energy analyzer (EA 125). Throughout the photoemission experiments were carried out at an average base pressure of ~$3.1 \times 10^{-11}$ torr with a power of 300 Watt. The total energy resolution, estimated from the width of the Fermi edge, was about 0.25 eV for the monochromatic AlKα line with photon energy of 1486.70 eV. The pass energy for the core level spectra was kept at 50 eV. Ar-ion sputtering performed at 2 KeV by maintaining an extractor pressure of 25 mPa. To get the estimation of change in the valence state of Eu, deconvolution was performed on all sputtered XPS spectra, using Peak Fit™ (v4, Jandel Scientific Software) on Eu ($3d_{5/2}$). Both Eu ($3d_{5/2}$) peaks were fitted with four Gaussian components, leading to the area under the curve. The precision of the peak fit analysis was better than 99% for both the compounds.

**Results and discussion**

**X-ray and EDX studies**

The Rietveld fit to the powder X-ray diffraction patterns of polycrystalline $Eu_2SrBi_2S_4F_4$ and $EuSr_2Bi_2S_4F_4$ is shown in Figure 1. Both samples were nearly phase pure and most of the peaks could be well indexed on the basis of the tetragonal structure (space group *I*4/*mmm*) of $Eu_3Bi_2S_4F_4$. Minor (<5%) impurity peaks corresponding to $EuF_{2.4}$ at 2θ = 26.9 degrees were present in $Eu_2SrBi_2S_4F_4$ and $EuSr_2Bi_2S_4F_4$. The same impurity phase has also been reported in $Eu_3Bi_2S_4F_4$[27]. The compositions of the two samples were examined by EDX spectroscopy, thereby confirming the intended ratio of Eu:Sr as 2:1 in $Eu_2SrBi_2S_4F_4$ and 1:2 in

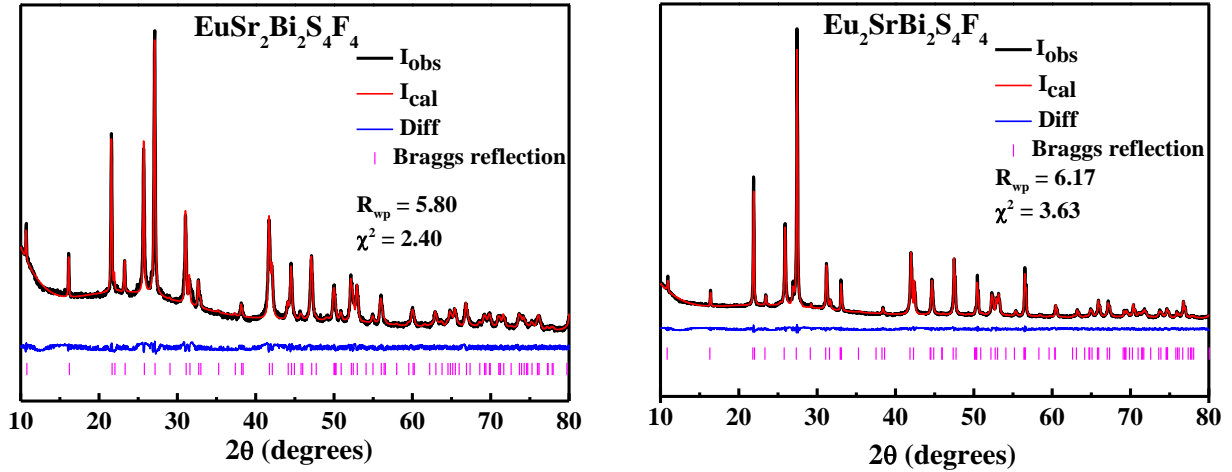

Figure 1(a). Rietveld fitted room temperature powder x-ray diffraction data of EuSr$_2$Bi$_2$S$_4$F$_4$ and Eu$_2$SrBi$_2$S$_4$F$_4$

EuSr$_2$Bi$_2$S$_4$F$_4$ respectively. The crystal structure of Eu$_{3-x}$Sr$_x$Bi$_2$S$_4$F$_4$ (x = 1 and 2), Z = 2, shown in Figure 1(b), consists of a Eu$_{3-x}$Sr$_x$F$_4$ (x = 1 and 2) spacer layers in place of Eu$_3$F$_4$ layers as in Eu$_3$Bi$_2$S$_4$F$_4$, alternating with square pyramidal BiS$_2$ bilayers.

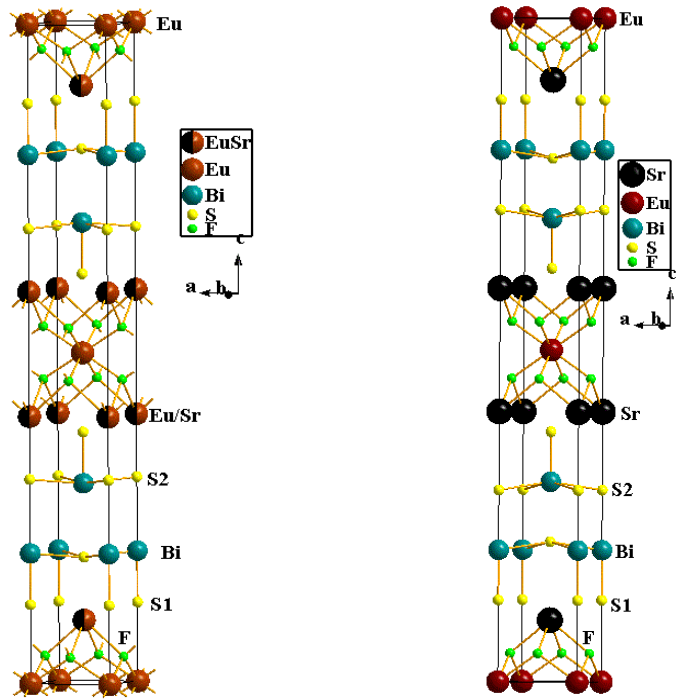

Figure 1(b). Crystal structure of Eu$_2$SrBi$_2$S$_4$F$_4$ and EuSr$_2$Bi$_2$S$_4$F$_4$

Structural parameters were calculated by Rietveld fitting of the powder x-ray diffraction data in the 2θ range 10°–80°, results of which are shown in Tables 1 and 2. In EuSr$_2$Bi$_2$S$_4$F$_4$ (Z =

2; two Eu-atoms and four Sr-atoms in a unit cell), both the Eu-atoms in the unit cell are assigned to fully occupy the 2$a$-sites and the four Sr-atoms are assigned to fill the 4$e$-sites completely. With this distribution, EuSr$_2$Bi$_2$S$_4$F$_4$ has a well ordered crystal structure. The structural refinement using this distribution of atoms resulted in a reasonably good $R_{wp}$ = 5.80% and a goodness-of-fit ($\chi^2$) = 2.40. In Eu$_2$SrBi$_2$S$_4$F$_4$, there are four Eu-atoms and two Sr-atoms in a unit cell. Out of the four Eu-atoms in a unit cell in this case, we assign two Eu-atoms to fully occupy 2$a$-sites. The 4$e$-sites in this case are assumed to be filled by two Eu-atoms and two Sr-atoms, with Eu- and Sr-atoms statistically distributed over the 4$e$-site in this material. With this distribution of atoms in the case of Eu$_2$SrBi$_2$S$_4$F$_4$, we could get an acceptable $R_{wp}$ = 6.17% with a g.o.f. ($\chi^2$) of 3.63.

Table1: Refined Structural parameter for EuSr$_2$Bi$_2$S$_4$F$_4$

| SG : $I4/mmm$; | a (Å) = 4.0628(1); | c (Å) = 32.805(1) |
|---|---|---|
| $R_{wp}$ = 5.80%; | $\chi^2$ = 2.40% | |

| Atom | Wyckoffsite | x | y | z | Occ(fixed) |
|---|---|---|---|---|---|
| Sr | 4$e$ | 0 | 0 | 0.5921(4) | 1 |
| Eu | 2$a$ | 0 | 0 | 0 | 1 |
| Bi | 4$e$ | 0 | 0 | 0.1979(3) | 1 |
| S | 4$e$ | 0 | 0 | 0.118(1) | 1 |
| S | 4$e$ | 0 | 0 | 0.705(1) | 1 |
| F | 8$g$ | 0 | 1/2 | 0.037(1) | 1 |

| | **Bond distance (Å)** | **Multiplicity** |
|---|---|---|
| Sr–F | 2.72(2) | 4 |
| Eu–F | 2.37(2) | 8 |
| Bi–S(1) | 2.62(3) | 1 |
| Bi–S(2) | 2.882(3) | 4 |

Table 2: Refined Structural parameterfor $Eu_2SrBi_2S_4F_4$

| SG: $I4/mmm$; | a (Å) = 4.0661(2); | c (Å) = 32.611(4) |
| --- | --- | --- |
| $R_{wp}$ = 6.17%; | $\chi^2$ = 3.63% | |

| Atom | Wyckoffsite | x | y | z | Occ (fixed) |
| --- | --- | --- | --- | --- | --- |
| Eu/Sr | 4e | 0 | 0 | 0.5945(5) | 0.5/0.5 |
| Eu | 2a | 0 | 0 | 0 | 1 |
| Bi | 4e | 0 | 0 | 0.1979(3) | 1 |
| S(1) | 4e | 0 | 0 | 0.120(1) | 1 |
| S(2) | 4e | 0 | 0 | 0.690(1) | 1 |
| F | 8g | 0 | 1/2 | 0.040(2) | 1 |

| | Bond distance (Å) | Multiplicity |
| --- | --- | --- |
| Eu/Sr–F | 2.70(2) | 4 |
| Eu–F | 2.42(2) | 8 |
| Bi–S(1) | 2.54(3) | 1 |
| Bi–S(2) | 2.887(3) | 4 |

The refined lattice parameters (as shown in Tables 1 and 2) are: For $Eu_2SrBi_2S_4F_4$ $a$ = 4.0661(2), $c$ = 32.611(4) Å and V = 539.16(8) Å$^3$; and for $EuSr_2Bi_2S_4F_4$ $a$ = 4.0628(1), $c$ = 32.805(1) Å and V = 541.47(3) Å$^3$. For the sake of comparison, we also give here the lattice parameters of the parent compound $Eu_3Bi_2S_4F_4$ $a$ = 4.0771(1), $c$ = 32.4330(6) Å and V = 539.1 Å$^3$ [27]. Comparing the ionic radii of $Sr^{2+}$ (1.26 Å) and $Eu^{2+}$ (1.25 Å) / $Eu^{3+}$ (1.07 Å) one would expect nearly the same lattice parameters for the three compounds[34].

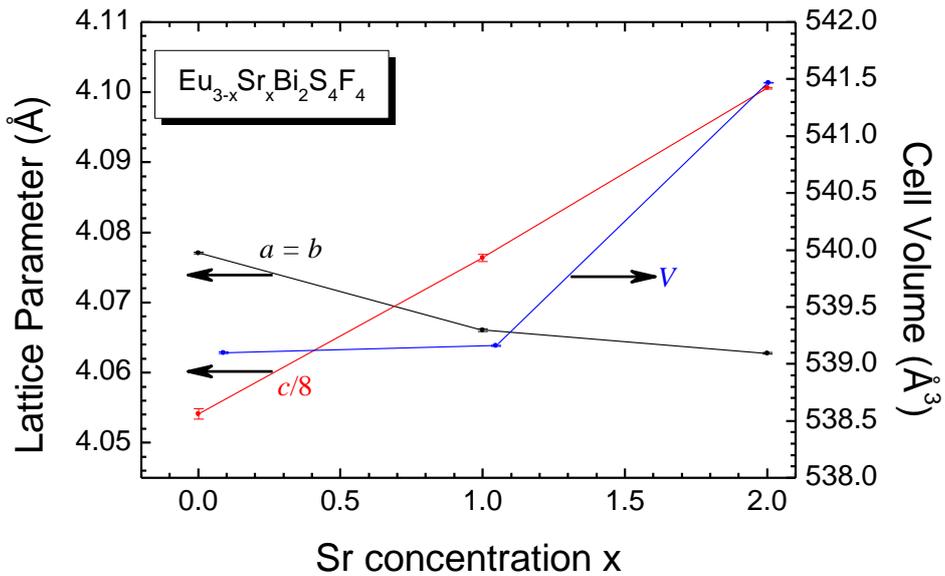

Figure 1(c). Variation of lattice parameter and Cell volume as function of Sr concentration x

Figure 1 (c) shows the variation of the lattice parameters with the Sr content *x*. It is evident that lattice parameter *a* decreases slightly with increasing Sr concentration, while the lattice parameter *c* and the cell volume V both increase with increasing Sr concentration. The Eu/Sr–F and Sr–F bond lengths in $Eu_2SrBi_2S_4F_4$ and $EuSr_2Bi_2S_4F_4$ are 2.70(4) and 2.72(2) Å, respectively. These bond lengths are quite similar to the Eu(1)–F bond length in $Eu_3Bi_2S_4F_4$ (2.678(5) Å). The other Eu(2)–F bonds in $Eu_2SrBi_2S_4F_4$, $EuSr_2Bi_2S_4F_4$ and $Eu_3Bi_2S_4F_4$ are 2.42(4), 2.37(2), and 2.403(6) Å, respectively. These bond lengths are smaller than the Eu(1)-F bond distances which clearly indicates a significant deviation of Eu from a divalent character. Thus from the structural analysis we assume that the Eu-atoms at the 2*a*-sites in both compounds are in a mixed valence state, closer to 3+ state. Earlier Zhai et al. have shown[27] that Eu-atoms occupying the 2*a*-sites in $Eu_3Bi_2S_4F_4$ are in mixed valence state. More precise information about the Eu-valence is obtained through our $^{151}$Eu Mössbauer spectroscopy (*vide infra*) and XPS studies.

**Mössbauer spectroscopic studies**

$^{151}$Eu Mössbauer spectra of the $Eu_2SrBi_2S_4F_4$ and $EuSr_2Bi_2S_4F_4$ samples are presented in Figures 2 and 3 along with transmission integral fits. The corresponding fitting parameters are listed in Table 3. Both samples show two well resolved signals that can be attributed to $Eu^{2+}$ (signals with an isomer shift around $\delta$ = –12.7 to –12.9 mm s$^{-1}$) and $Eu^{3+}$ (signals with an isomer shift around $\delta$ = 0.1 to 0.2 mm s$^{-1}$). These values of the isomer shifts compare well with those of ionically bonded di- and tri-valent europium. The $Eu^{2+}$ signals show some degree of quadrupole splitting ($\Delta E_Q$), a consequence of the non-cubic site symmetry. The experimental line width values ($\varGamma$) are slightly higher than usually observed (around 2.3 mm s$^{-1}$).

**(a) EuSr$_2$Bi$_2$S$_4$F$_4$**

In Eu$_3$Bi$_2$S$_4$F$_4$, the bond valence sum (BVS) calculations[27] suggest a divalent state of the Eu-atoms occupying the 4*e*-sites and a strongly mixed valence state (~60% trivalent component) of Eu-atoms occupying the 2*a*-sites. Considering that the Eu-bond lengths in EuSr$_2$Bi$_2$S$_4$F$_4$ and Eu$_2$SrBi$_2$S$_4$F$_4$ are close to those in Eu$_3$Bi$_2$S$_4$F$_4$, similar valence states of Eu-atoms at the 2*a*- and 4*e*-sites in our compounds may be expected.

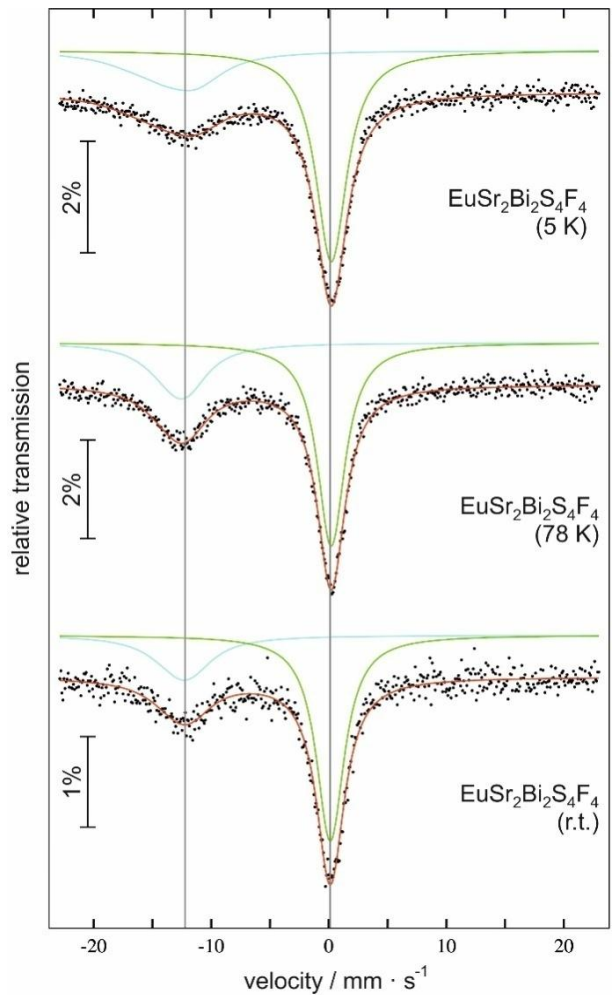

Figure 2. Experimental (data points) and simulated (continuous lines) $^{151}$Eu Mössbauer spectra of EuSr$_2$Bi$_2$S$_4$F$_4$ at 5K, 78 K and ambient temperature (r.t.).

At room temperature, the intensity of the signal corresponding to Eu$^{2+}$ is about 30% of that of Eu$^{3+}$ as shown in figure 2. This is close to the theoretical Eu$^{2+}$:Eu$^{3+}$ ratio on the basis of the BVS calculations (as pointed out above) considering that the Eu-atoms occupy only the 2*a*- sites. This also means that *we see similar Eu-mixed valence in* EuSr$_2$Bi$_2$S$_4$F$_4$ *as was observed in EuFBiS$_2$*[26]. We also note that at 78 and 5 K the relative intensities of the Eu$^{2+}$: Eu$^{3+}$ lines remain nearly same as those of the two lines at room temperature (see Table 3).

Table 3: Fitting parameters of $^{151}$Eu Mössbauer spectroscopic measurements at 5 K, 78 K and ambient temperature. $\delta$ = isomer shift, $\Delta E_Q$ = electric quadrupole splitting, $\Gamma$ = experimental line width. Parameters marked with an asterisk were kept fixed during the fitting procedure.

| Compound | $\delta$ (mm·s$^{-1}$) | $\Delta E_Q$ (mm·s$^{-1}$) | $\Gamma$ (mm·s$^{-1}$) | ratio |
|---|---|---|---|---|
| EuSr$_2$Bi$_2$S$_4$F$_4$ (5 K) | -13.1(1) | 10(1) | 5.7(5) | 31(1) |
| | 0.23(1) | 0* | 3.21(4) | 69(1) |
| EuSr$_2$Bi$_2$S$_4$F$_4$ (78 K) | −12.68(7) | 4(1) | 4.5(4) | 29(1) |
| | 0.21(1) | 0* | 2.86(4) | 71(1) |
| EuSr$_2$Bi$_2$S$_4$F$_4$ (r.t.) | −12.4(1) | 3(3) | 4.9(8) | 29(2) |
| | 0.13(2) | 0* | 2.87(5) | 71(1) |
| Eu$_2$SrBi$_2$S$_4$F$_4$ (5 K) | -12.68(3) | 5.9(2) | 3.2(1) | 53(1) |
| | 0.27(1) | 0* | 3.00(4) | 47(1) |
| Eu$_2$SrBi$_2$S$_4$F$_4$ (78 K) | −12.72(2) | 5.4(2) | 2.98(8) | 54(1) |
| | 0.27(1) | 0* | 2.89(4) | 46(1) |
| Eu$_2$SrBi$_2$S$_4$F$_4$ (r.t.) | −12.76(4) | 4.9(4) | 2.9(2) | 48(1) |
| | 0.16(2) | 0* | 2.86(6) | 52(1) |

**(b) Eu$_2$SrBi$_2$S$_4$F$_4$**

The Mössbauer spectrum of Eu$_2$SrBi$_2$S$_4$F$_4$ in figure 3 presents yet another interesting feature of the distribution of Eu-atoms in these compounds. This spectrum exhibits two Eu-lines with isomer shifts well known for Eu$^{2+}$ and Eu$^{3+}$ states (as shown in Table 3). The areas under the peaks corresponding to the Eu$^{2+}$ and Eu$^{3+}$ Mössbauer lines at room temperature are in the ratio of 48/52 ~ close to 1 which is consistent with all the Eu-atoms at the 4*e*-sites as in a stable valence state of Eu$^{2+}$ and the Eu-atoms at the 2*a*-sites are in nearly Eu$^{3+}$-state just as in EuSr$_2$Bi$_2$S$_4$F$_4$. At 78 and 5 K, the intensity of the Eu$^{2+}$-line is found to increase slightly (see figure 3 and Table 3). Since the Eu-atoms at the 4*e*-sites are in a Eu$^{2+}$-state already, this weak enhancement of intensity of Eu$^{2+}$-line shows that Eu-atoms at the 2*a* sites acquire a small 2+- character at 78 and at 5 K. This establishes weakly temperature dependent mixed valence character of Eu-atoms at the 2*a*-sites.

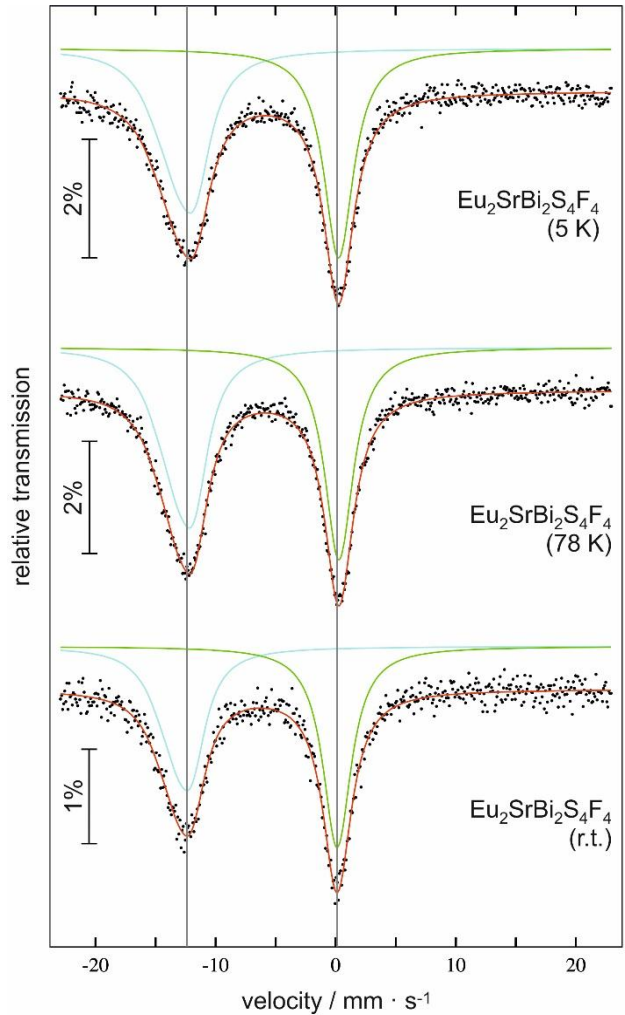

Figure 3. Experimental (data points) and simulated (continuous lines) $^{151}$Eu Mössbauer spectra of Eu$_2$SrBi$_2$S$_4$F$_4$ at 5 K, 78K and ambient temperature (r.t.).

To summarize this discussion of the Mössbauer spectroscopic measurements: In EuSr$_2$Bi$_2$S$_4$F$_4$, *all the Eu-atoms* are located at the same crystallographic 2*a*-sites. Thus observing *two signals in the Mössbauer spectra* corresponding to Eu$^{2+}$ and Eu$^{3+}$ states clearly suggest that they are *in a homogeneous mixed valence state*. In Eu$_2$SrBi$_2$S$_4$F$_4$, the situation is somewhat different. Eu-atoms in this case are located at both 2*a* and 4*e* sites. The Eu-atoms at the 4*e*-sites are in stable valence state (2+) and the Eu-atoms at the 2*a*-sites are in a mixed valence state but close to 3+ at room temperature. At 78 and 5 K, there is a perceptible enhancement of the intensity of the $Eu^{2+}$-signal, implying that the Eu-valence shifts towards 2. The decrease of the valency of Eu (approach towards $Eu^{2+}$) with decreasing the temperature, as observed in the present work, is in contrast with the situation in well known intermetallic Eu-mixed valence materials, as for example $EuCu_2Si_2$[35] and $EuPd_2Si_2$[36], wherein the valence of Eu increases strongly towards 3 with the *decrease* of temperature. In this sense, the temperature variation of the average valence observed in this work is rather unusual.

We must point out here that in homogeneous intermetallic mixed valence materials such as $EuPd_2Si_2$, the two participating valence states $Eu^{2+}$ and $Eu^{3+}$ in the valence fluctuations phenomenon are not observed separately in $^{151}$Eu-Mössbauer spectroscopy. This is because the Eu-Mössbauer characteristic time $\tau_{MOSS}$ is much larger than the valence fluctuation time $\tau_{VF}$ in intermetallics and therefore Eu-Mössbauer spectroscopy sees an *average line*. As in $EuSr_2Bi_2S_4F_4$, $Eu_2SrBi_2S_4F_4$ and in $EuFBiS_2$, two distinct $Eu^{2+}$ and $Eu^{3+}$ lines are observed, it means that in these materials, $\tau_{VF} \gg \tau_{MOSS}$. To the best of our knowledge, besides our two materials $EuSr_2Bi_2S_4F_4$ and $Eu_2SrBi_2S_4F_4$, only two other such materials $EuFBiS_2$ and $Eu_3Bi_2S_4F_4$ are known with $\tau_{VF} \gg \tau_{MOSS}$.

**X-ray Photoemission Spectroscopy (XPS) studies**

Systematic X-ray photoemission spectroscopy experiments have been carried out for $Eu_{3-x}Sr_xBi_2S_4F_4$ with x = 1 and 2. The experimental conditions were kept identical for the experiments. During the photoemission studies, small specimen charging was observed which was later calibrated by assigning the C 1s signal at 284.6 eV. To get a clean surface, Ar ion sputtering was performed on the samples. Figure 4 shows the x-ray photoelectron 3d core-level spectra of Eu in $Eu_2SrBi_2S_4F_4$ and $EuSr_2Bi_2S_4F_4$. The main features of the spectra are the 3d spin-orbit split states $3d^{2+}_{3/2}$, $3d^{2+}_{5/2}$, $3d^{3+}_{3/2}$ and $3d^{3+}_{5/2}$ at binding energy 1155, 1125, 1165 and 1135 eV respectively. We also see a relatively much weaker satellite peak (marked as "sat") at ~ 8 eV higher binding energy (B.E.) relative to the B.E. of the peak $3d^{3+}_{5/2}$. This *sat* peak originates from the multi-electronic excitations during the photoelectron emission process as has been observed in a number of Eu-compounds ,e.g. $EuF_3$, $EuCl_3$ and $Eu_2O_3$[37, 38]. As we are not concerned with such effects in this work, we shall not talk about this satellite peak any more.

Qualitatively speaking, intensities of the four spin-orbit split states contain information *about the mixed valence state of the Eu-ions in these compounds and also about the surface states*. The area under the curve referred as % change in the valence state of Eu ($3d_{5/2}$) is obtained (from peak deconvolution, not shown here) and listed in Table 4. We first look at the XPS spectrum of $EuSr_2Bi_2S_4F_4$. As described above, Eu-atoms in this material occupy only the 2*a*-sites. Thus, just as in the Mössbauer spectroscopic studies, our observation of the two sets of peaks, $Eu^{2+}$ peaks ($3d^{2+}_{3/2}$ and $3d^{2+}_{5/2}$) *and* $Eu^{3+}$ peaks ($3d^{3+}_{3/2}$ and $3d^{3+}_{5/2}$ ) demonstrates clearly and independently occurrence of the mixed valence phenomenon associated with Eu-ions in this material. What is new in the XPS spectra, however, is the evidence of the presence of the *enhanced* $Eu^{2+}$-states in the surface layer of the sample. Consider, for example, the ratio of the intensities of the two XPS lines $3d^{2+}_{5/2}$ and $3d^{3+}_{5/2}$. A simple estimate shows that this ratio is ~ 43/57 (Figure 4, Table 4), much bigger than 0.29/0.71, the ratio of the populations of the two valence states $Eu^{2+}$ and $Eu^{3+}$, as inferred from the Mössbauer spectroscopic data, in the mixed valence state of Eu-atoms in the bulk of the material. At this point, it must be stressed that XPS is a surface sensitive technique and explores only about 100 Å thick surface layer. Thus the extra population of $Eu^{2+}$-states as detected by XPS must be due to the surface layer having a larger population of $Eu^{2+}$-states. Such a transition from $Eu^{3+}$-states in the bulk to $Eu^{2+}$-state in the surface layer has been reported for several Eu-compounds such as $EuF_3$, $EuCl_3$ and $Eu_2O_3$[37].

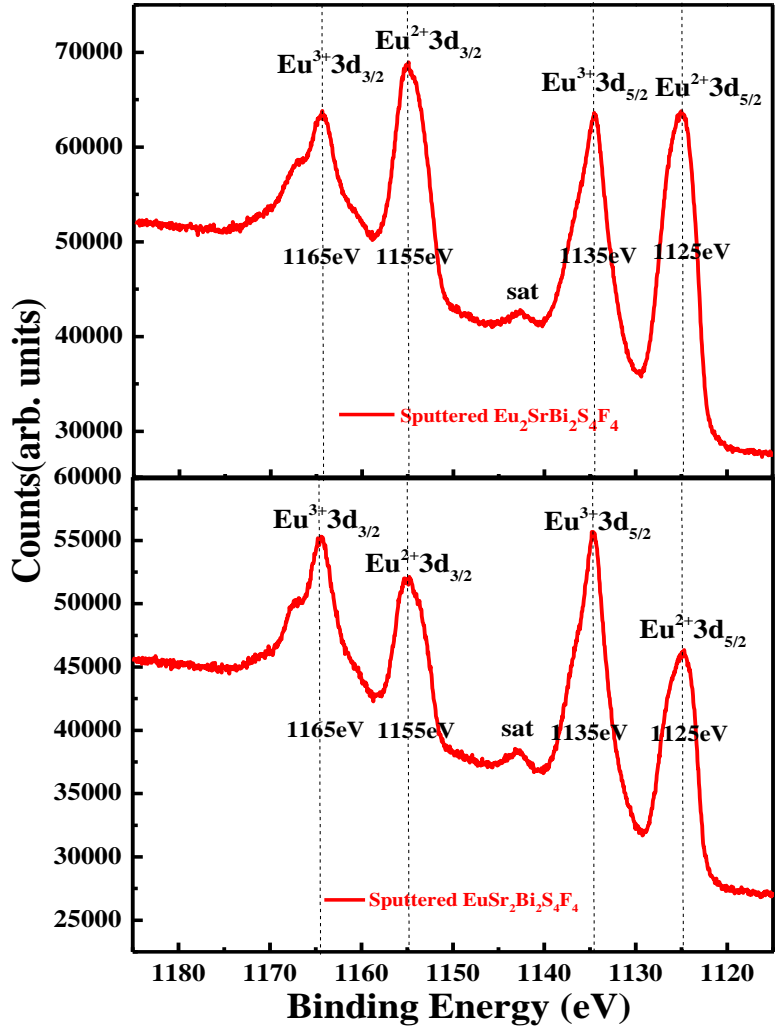

Figure 4. XPS core-level spectra of Eu (3d) in $Eu_{3-x}Sr_xBi_2S_4F_4$ (x = 1 and 2) at 300K.

Table 4: Change in the Valence state of Eu obtained as % Area under the curve for $Eu_{3-x}Sr_xBi_2S_4F_4$

| x | $Eu^{2+}$ ($3d_{5/2}$) | $Eu^{3+}$ ($3d_{5/2}$) |
|---|---|---|
| 1 | 54.8% | 45.2% |
| 2 | 42.7% | 57.3% |

In the case of $Eu_2SrBi_2S_4F_4$ also we observe two sets of peaks, $Eu^{2+}$ states ($3d^{2+}_{3/2}$ and $3d^{2+}_{5/2}$) *and* $Eu^{3+}$ states ($3d^{3+}_{3/2}$ and $3d^{3+}_{5/2}$). At room temperature, the valence state of Eu-atoms occupying the 2*a*-sites is very nearly 3+ as we discussed above while considering the Mössbauer spectra of the compound. An equal number of Eu-atoms, occupying the 4*e*-sites, are in a $Eu^{2+}$-state. Thus we should see nearly the same intensity of these two sets of Eu-XPS peaks. The intensity ratio of the two XPS peaks of $3d^{2+}_{5/2}$ and $3d^{3+}_{5/2}$ (~ 55/45) (Figure 4, Table 4) clearly shows that the $Eu^{2+}$ states have higher population than that of $Eu^{3+}$ states. This higher intensity ratio $Eu^{2+}/Eu^{3+}$ of the XPS-peaks in $Eu_2SrBi_2S_4F_4$ also clearly shows the contribution of the $Eu^{2+}$ states originating from the surface layer.

**Magnetization studies**

A most remarkable effect that we observe in these materials is a strong variation of the Eu-valence by the application of a magnetic field as we discuss below. We must stress here that we have carried out the magnetization measurements on two independent samples and obtained very similar results. We show in Figure 5 results of our magnetization measurements at 5 K in $EuSr_2Bi_2S_4F_4$ as well as in $Eu_2SrBi_2S_4F_4$ as a function of applied magnetic field. The saturation magnetic moment at 5 K in $EuSr_2Bi_2S_4F_4$ in an applied field of 9 T is 1.75 $\mu_B$ / f.u. From this value of Eu-moment (there are no other magnetic species in the material), we estimate the fraction of $Eu^{2+}$-state as 1.75/7 = 25% which corresponds to an average valence of Eu as 2.75. This estimate of the fractional populations of the two valence states of Eu is consistent with the intensities of the two lines in the $^{151}$Eu-Mössbauer spectrum of $EuSr_2Bi_2S_4F_4$, shown in figure 2.

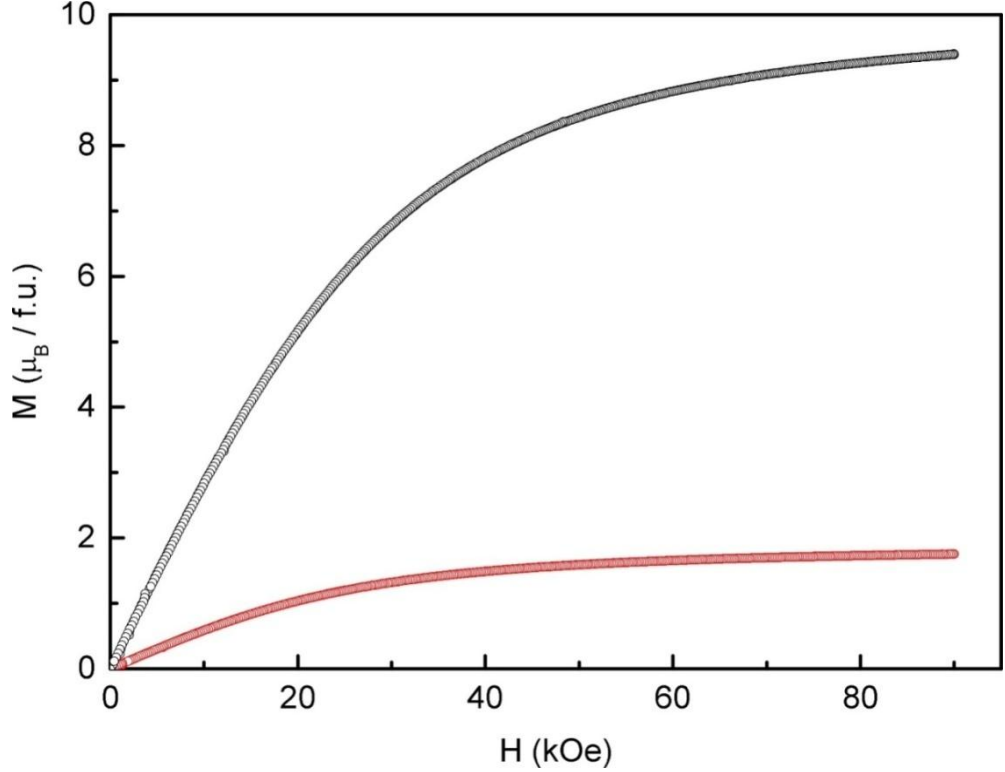

Figure 5. Magnetization isotherms of $Eu_2SrBi_2S_4F_4$ (black) and $EuSr_2Bi_2S_4F_4$ (red) at 5 K

There are already examples of Eu- and Yb-valence changes in applied magnetic fields reported in literature[39–41]. We find evidence of a similar valence change in an applied field in the case of $Eu_2SrBi_2S_4F_4$. The measured saturation magnetic moment is 9.39 $\mu_B$ / f.u. A large fraction (7 $\mu_B$) of this saturation moment arises from the divalent Eu-atoms at the 4*e*-sites. The remaining contribution of 2.39 $\mu_B$ must come from the mixed valent Eu-atoms at the 2*a*-sites, which means that in the applied field the mixed valence of Eu-atoms at the 2*a*-sites is 2.39/7 ~ 2.65. The Mössbauer spectrum of $Eu_2SrBi_2S_4F_4$, figure 3, shows both the $Eu^{2+}$- and $Eu^{3+}$-lines with nearly equal integrated intensities. This implies that Eu-atoms occupying 2*a*-sites in this material which are nearly 3+ in zero applied field move towards 2+ state in the presence of a magnetic field. Mitsuda et al. observed similar effects, namely field induced valence change, in applied fields of ~ 100 T in $Eu(Pd_{1-x}Pt_x)_2Si_2$[39]. To put it concisely, in $Eu_2SrBi_2S_4F_4$ we observe the *remarkable phenomenon of a field-induced change of the Eu-valence, moving from $Eu^{3+}$ to $Eu^{2+}$ in the applied magnetic field.* Non-availability of a high field option in our Mössbauer spectrometer does not permit us to observe this remarkable phenomenon directly, namely, a growth of the intensity of the $Eu^{2+}$-line as a function of applied magnetic field.

**Resistivity studies**

Results of resistivity ρ (T) measurements at ambient pressure of $Eu_2SrBi_2S_4F_4$ and $EuSr_2Bi_2S_4F_4$ are shown in Figure 6. Both samples show a semi-metallic behavior in the temperature range of 2 – 300 K. At room temperature, the resistivity of both samples is quite low; ~ 1-1.5 mΩcm for $Eu_2SrBi_2S_4F_4$ and $EuSr_2Bi_2S_4F_4$. In the former compound, ρ (T) remains more or less temperature independent down to 2 K whereas in the latter, it rises to ~ 3 mΩcm at low temperature. The temperature dependence and these values of the resistivity are comparable to those of $Eu_3Bi_2S_4F_4$ which is known to exhibit superconductivity below $T_c$ ~ 1.5 K[27]. We do not observe superconductivity in $Eu_2SrBi_2S_4F_4$ and $EuSr_2Bi_2S_4F_4$ down to 2 K. However, the possibility of superconductivity below 2 K in these materials cannot be ruled out.

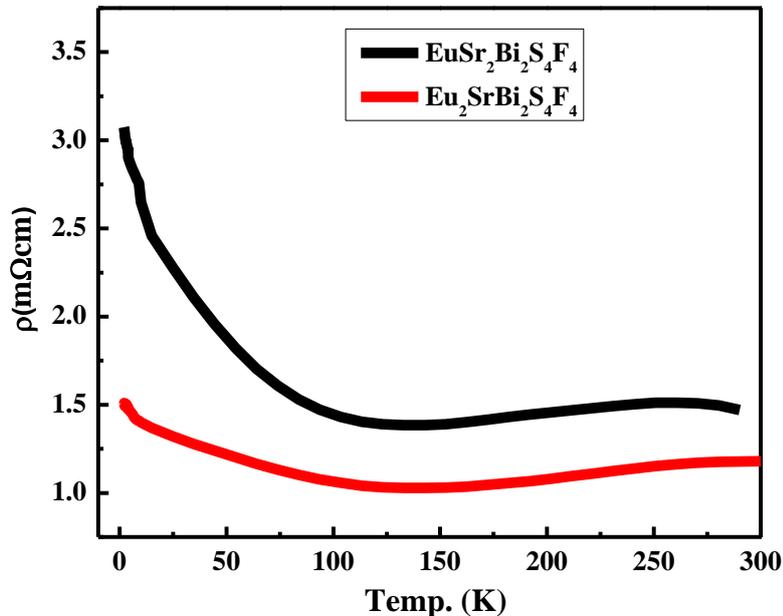

Figure 6. Variable temperature resistivity at ambient pressure from 2–300 K

## Conclusion

We have successfully synthesized the polycrystalline compounds $Eu_2SrBi_2S_4F_4$ and $EuSr_2Bi_2S_4F_4$, which are new derivatives of the Eu-3244 and crystallize in a tetragonal structure ($I4/mmm$). In $EuSr_2Bi_2S_4F_4$, Eu-atoms occupy the 2$a$-sites and Sr-atoms occupy the 4$e$-sites. In $Eu_2SrBi_2S_4F_4$, 2$a$- sites are fully occupied by Eu-atoms and the 4$e$-sites are occupied by Eu-atoms and Sr-atoms in equal proportion and are randomly distributed. Eu-atoms at the 2$a$-sites in both $EuSr_2Bi_2S_4F_4$ and $Eu_2SrBi_2S_4F_4$ are in a homogeneous mixed valence state with fluctuations of the Eu-valence *slow* with respect to the Mössbauer spectroscopy time scale. XPS studies also reveal the occurrence of Eu-valence fluctuation. In $EuSr_2Bi_2S_4F_4$ we find the average Eu valence to be ~ 2.7 from Mössbauer spectroscopy and magnetization experiments as well. An important observation of our magnetization results of $Eu_2SrBi_2S_4F_4$ is that the valence state of Eu at 2$a$-sites moves from nearly 3+ (in zero applied field) to the estimated valence of 2.65 in the presence of applied magnetic field. This is a significant field induced valence change in fields of the order of 10 T.

## Acknowledgements

Authors thank DST for the PPMS and National SQUID facility at the department of Physics, IIT Delhi. AKG thanks DST for providing financial support. ZH and GST thank UGC and CSIR, respectively, for a fellowship. GKS and SA acknowledge the department of science and technology (SERB & TSDP), CEFIPRA, New Delhi, UGC (BSR-RFSMS-SRF-Meritorious fellowship, SAP, MRP) for their financial support.

[†]Visiting scientist at Solid State and Nano Research Laboratory, Department of Chemistry, IIT Delhi